\documentclass[aps,prl,twocolumn,amsmath,amssymb]{revtex4-1}

\usepackage{amsmath}
\usepackage{amssymb}

\usepackage[pdftex]{color}
\usepackage{graphicx}
\usepackage{dcolumn} 
\usepackage{bm} 
\usepackage{epic}
\usepackage{hyperref}
\usepackage{longtable}
\usepackage{ulem}   
\newcommand{\bs}[1]{{\boldsymbol{#1}}}

\newcommand{\HH}{\mathcal{H}}
\newcommand{\kk}{\boldsymbol{k}}
\newcommand{\T}{\boldsymbol{T}}

\begin{document}
\title{
Chiral $d$-wave Superconductivity in SrPtAs}

\author{Mark H. Fischer${}^{1,2}$}
\author{Titus Neupert${}^{3,4}$}
\author{Christian Platt${}^{5}$}
\author{Andreas P. Schnyder${}^{6}$}
\author{Werner Hanke${}^{5}$}
\author{Jun Goryo${}^{7}$}
\author{Ronny Thomale${}^{5}$}
\author{Manfred Sigrist${}^{4}$}

\affiliation{${}^{1}$
Department of Condensed Matter Physics, Weizmann Institute of Science, Rehovot 76100, Israel}
\affiliation{${}^{2}$
Department of Physics, Cornell University, Ithaca, New York 14853, USA}
\affiliation{${}^{3}$
Princeton Center for Theoretical Science, Princeton University, Princeton, New Jersey 08544, USA}
\affiliation{${}^{4}$
Institut f\"ur Theoretische Physik, ETH Z\"urich, CH-8093 Z\"urich, Switzerland}
\affiliation{${}^{5}$
Institute for Theoretical Physics, University of W\"urzburg, D-97074 W\"urzburg, Germany}
\affiliation{${}^{6}$
Max-Planck-Institut f\"ur Festk\"orperforschung, Heisenbergstrasse 1, D-70569 Stuttgart, Germany}
\affiliation{${}^{7}$
Department of Advanced Physics, Hirosaki University, Hirosaki 036-8561, Japan}

\date{\today}

\begin{abstract}
Recent $\mu$SR measurements on SrPtAs revealed time-reversal-symmetry
breaking with the onset of superconductivity [Biswas {\it et al.},
Phys. Rev. B {\bf 87}, 180503(R) (2013)], suggesting an unconventional
superconducting state.
We investigate this possibility via functional renormalization group and find a chiral $(d+\mathrm{i}d)$-wave order parameter favored by the
multiband fermiology and hexagonal symmetry of SrPtAs. 
This $(d+\mathrm{i}d)$-wave state exhibits
significant gap anisotropies as well as gap differences on the different
bands, but only has
point nodes on one of the bands at the Brillouin zone corners.
We study the topological characteristics of this superconducting phase, which 
features Majorana-Weyl nodes in the bulk, 
protected surface states, and an associated thermal Hall response.
The lack of extended nodes and the spontaneously broken time-reversal symmetry  of the $(d+\mathrm{i}d)$-wave state are in agreement with the $\mu$SR experiments. 
Our theoretical findings together with the experimental evidence thus suggests that SrPtAs is the first example of chiral $d$-wave superconductivity.
\end{abstract}

\maketitle

\emph{Introduction.}
SrPtAs is a pnictide superconductor ($T_c=2.4$~K), where present
experimental evidence strongly suggests broken time-reversal symmetry (TRS) in the superconducting state~\cite{nishikubo:2011,
  biswas:2013}.
So far, this property has only been found in nature for a limited number
of compounds such as $\text{Sr}_2\text{RuO}_4$~\cite{maeno-nature,maeno-nature-2}, like SrPtAs a quasi-two-dimensional material, and naturally offers the possibility of a chiral superconducting state with non-trivial topological properties~\cite{RevModPhys.63.239}. Unlike other pnictide superconductors, the crystal structure of SrPtAs has hexagonal symmetry. 
This has important consequences for possible chiral superconducting states: While square-lattice symmetry as found in $\text{Sr}_2\text{RuO}_4$ generically triggers a
chiral $p$-wave triplet order parameter due to the degeneracy of $p_x$-
and $p_y$-wave at the instability level, hexagonal symmetry implies degeneracy in the $d_{x^2-y^2}$- and $d_{xy}$-wave channel, too.
In such a case, similar to the
$p$-wave scenario on a square lattice, a chiral ($d_{x^2-y^2}+\mathrm{i} d_{xy}$)-wave order-parameter combination that spontaneously breaks TRS
generically maximizes the condensation energy of the
superconducting state~\footnote{For the hexagonal system, this line of
reasoning is not restricted to the Cooper channel, but likewise
applies to the particle-hole channel~\cite{maharaj:2013}.}.

Chiral superconductors exhibit many exotic phenomena due to their nontrivial topology~\cite{RevModPhys.63.239,PhysRevB.60.4245,PhysRevLett.85.4940}, such as 
Majorana vortex bound states and gapless chiral edge modes, that carry quantized thermal or spin currents.
Chiral $d$-wave superconductivity has previously been proposed in various model
calculations for graphene doped to van Hove filling~\cite{PhysRevB.78.205431,honerkamp:2008,nandkishore:2012,kiesel:2012,PhysRevB.85.035414}
and has recently been propagated to explain the superconducting state in water-intercalated sodium
cobaltates~\cite{PhysRevLett.111.097001}. Note, however, that there exists a natural competition between $d$-wave and $f$-wave superconductivity in these scenarios, a recurrent motif in the study of superconducting
instabilities in hexagonal systems~\cite{platt:2013tmp}.
This is intuitively illustrated by a single-orbital honeycomb Hubbard
model: While the leading instability is of $d$-wave symmetry for bands
close to the van-Hove singularity, it changes to $f$-wave symmetry
when the Fermi surface consists of (disconnected) pockets around the Brillouin zone (BZ) corners. The $f$-wave instability is preferred in that case, 
because all gap nodes can be placed such that they do not
intersect with the Fermi surface~\cite{raghu:2010b, kiesel:2012}.
So far, no
unambiguous experimental evidence in support of chiral $d$-wave
superconductivity in a hexagonal system exists, hence awaiting further investigation and
refinement.

\begin{figure}[tb]
  \centering
  \includegraphics[width=8cm]{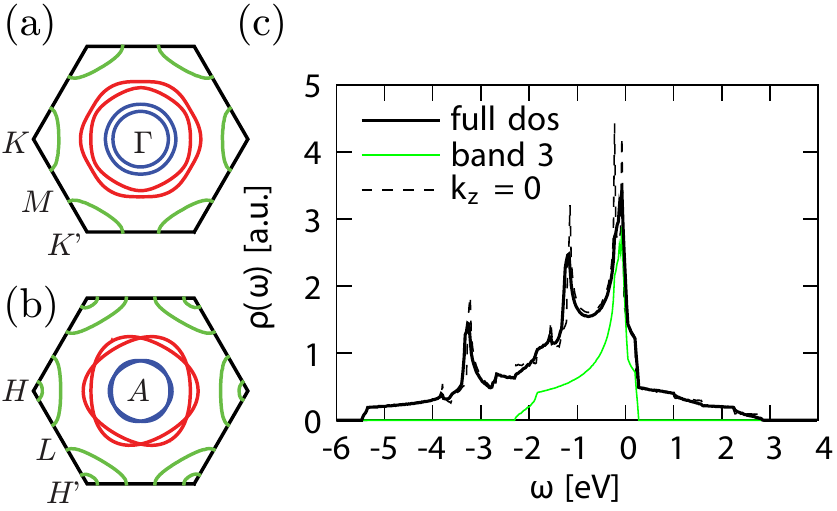}
  \caption{Fermi surface at $k_z=0$ (a) and $k_z=\pi$ (b) of SrPtAs. (c) The density of states for the full three-dimensional bandstructure (bold black), for a (2D) plane with $k_z=0$ (dashed black), and the three-dimensional bandstructure only looking at bands around the corners of the Brillouin zone (band $3$, green line). Note the van Hove singularity very close to the Fermi energy.}
  \label{fig:bs}
\end{figure}

In this Letter, we present functional renormalization group (FRG)~\cite{RevModPhys.84.299,platt:2013tmp}
studies for SrPtAs, which, in combination with the experimental evidence at hand,
render this system a prime candidate for chiral $d$-wave
superconductivity.
So far, superconductivity in SrPtAs has only been investigated within a mean-field approach with a generic short-range density-density interaction~\cite{goryo:2012}.
Depending on the specific structure of the interaction the leading instability is either in the $s$- or $f$-wave channel, with an additional $d$-wave solution close by in energy, however.
Our analysis sheds further light on the
nature of superconductivity in SrPtAs and allows for a substantiated microscopic perspective.
For intermediate interactions, we find $d$-wave superconductivity as the
dominant Fermi surface instability, while
ferromagnetism is non-negligible due to the large, partly unnested
density of states at the Fermi level.
This trend towards $d$-wave superconductivity can be attributed to
the multiband fermiology of SrPtAs shown in Fig.~\ref{fig:bs}(a). While the 
pockets centered around $K$ and $K'$ in the BZ are the main driver for superconductivity, 
proximity-coupled pockets around the $\Gamma$ point are a crucial ingredient to tilt the system in favor of a $d$-wave instability. 

In a two-dimensional (2D) geometry, the chiral $d$-wave state is a
topological superconductor that supports dispersing chiral Majorana
modes at the edge. These edge modes are robust to perturbations, for
they are protected by the emergent particle-hole symmetry of the superconducting condensate.
We elaborate in this Letter that the fate of this topological state
in the three-dimensional (3D) material SrPtAs is a Weyl superconducting
state~\cite{meng:2012}. As such, it features Majorana surface states and nodal points
in the bulk which are described as three-dimensional Majorana-Weyl
fermions in the low-energy limit. 

\emph{Model.}
The low-energy electronic structure of SrPtAs can be described by three bands, indexed by $b=1,2,3$, stemming from the Pt $d$ orbitals with dispersions~\cite{youn:2012}
\begin{equation}
  \xi_{\kk\pm}^{(b)} = \epsilon^{(b)}_{1\kk} - \mu^{(b)} \pm \sqrt{|\epsilon^{(b)}_{c\kk}|^2 + (\alpha^{(b)}_{\kk})^2}
  \label{eq:bands}.
\end{equation}
For each band, the individual terms in Eq.~\eqref{eq:bands} are given by $\epsilon^{(b)}_{1\kk}=t^{(b)}\sum_{n}\cos(\T_n\cdot \kk) + t^{\prime(b)}_z \cos k_z$, $\epsilon^{(b)}_{c\kk} = t^{(b)}_z \cos(\frac{ k_z}2)[1 + e^{- \mathrm{i} \T_3\cdot\kk} + e^{\mathrm{i}\T_2\cdot\kk}]$, 
and the spin-orbit coupling $\alpha^{(b)}_{\kk}= \alpha^{(b)} \sum_{n}\sin(\T_n\cdot \kk)$.
The lattice vectors are $\T_1 = (0, 1, 0)$, $\T_2 = (\sqrt{3}/2, -1/2,0)$, $\T_3 =-\T_1-\T_2$, and all the lattice constants have been set to unity. 
Figure~\ref{fig:bs} shows the Fermi surfaces at $k_z=0$ (a) and $k_z=\pi$ (b) obtained with the tight-binding parameters from Ref.~\cite{youn:2012} and reflects the quasi-two-dimensional nature of this material. 
Note that the outermost band is close to the van-Hove singularity (located at $M)$ and hence contributes most to the density of states (DOS), namely 74$\%$ of the total DOS, see Fig.~\ref{fig:bs}(c).

While spin-orbit coupling and the $k_z$ dispersion in this compound play a crucial role for the mixing of order parameters~\cite{fischer:2011b, goryo:2012} and magnetic properties~\cite{youn:2012}, the dominant instability likely depends more on the geometry of the Fermi surfaces and the low-energy spectrum. We therefore focus on the five bands crossing the Fermi energy at $k_z=0$~\footnote{Our results do not change upon choosing $k_z=\pi$}  with dispersions given in Eq.~\eqref{eq:bands} and treat them each as (spin-degenerate) independent bands with operators $\psi_{\beta\kk s}$ ($\beta = 1 \dots 5$) in a two-dimensional band structure. We then introduce the interaction Hamiltonian
\begin{equation}
\begin{split}
  \HH' =&\,\hphantom{+\,} G_1 \sum_{\beta< \beta'}\sum_{\kk,s}\psi^{\dag}_{\beta\kk_1 s}\psi^{\dag}_{\beta'\kk_2 s'}\psi^{\phantom{\dag}}_{\beta'\kk_3 s'}\psi^{\phantom{\dag}}_{\beta\kk_4 s}
  \\
  & + G_2 \sum_{\beta< \beta'}\sum_{\kk,s}\psi^{\dag}_{\beta\kk_1 s}\psi^{\dag}_{\beta'\kk_2 s'}\psi^{\phantom{\dag}}_{\beta\kk_3 s'}\psi^{\phantom{\dag}}_{\beta'\kk_4 s}
  \\
  & + G_3 \sum_{\beta<\beta'}\sum_{\kk,s}\psi^{\dag}_{\beta\kk_1 s}\psi^{\dag}_{\beta\kk_2 s'}\psi^{\phantom{\dag}}_{\beta'\kk_3 s'}\psi^{\phantom{\dag}}_{\beta'\kk_4 s} 
  \\
  & + G_4 \sum_{\beta}\sum_{\kk,s}\psi^{\dag}_{\beta\kk_1 s}\psi^{\dag}_{\beta\kk_2 s'}\psi^{\phantom{\dag}}_{\beta\kk_3 s'}\psi^{\phantom{\dag}}_{\beta\kk_4 s}
  \label{eq:Hint}
  \end{split}
\end{equation}
containing inter-band ($G_1$) and intra-band ($G_4$) density-density interactions, an exchange interaction ($G_{2}$), and a pair-hopping term ($G_3$). Note that the sum $\sum_{\kk,s}$ runs over all spins and momenta constrained to $\kk_1 + \kk_2 = \kk_3 + \kk_4$ (modulo reciprocal lattice vectors).

\begin{figure}[t]
  \centering
  \includegraphics[width=8cm]{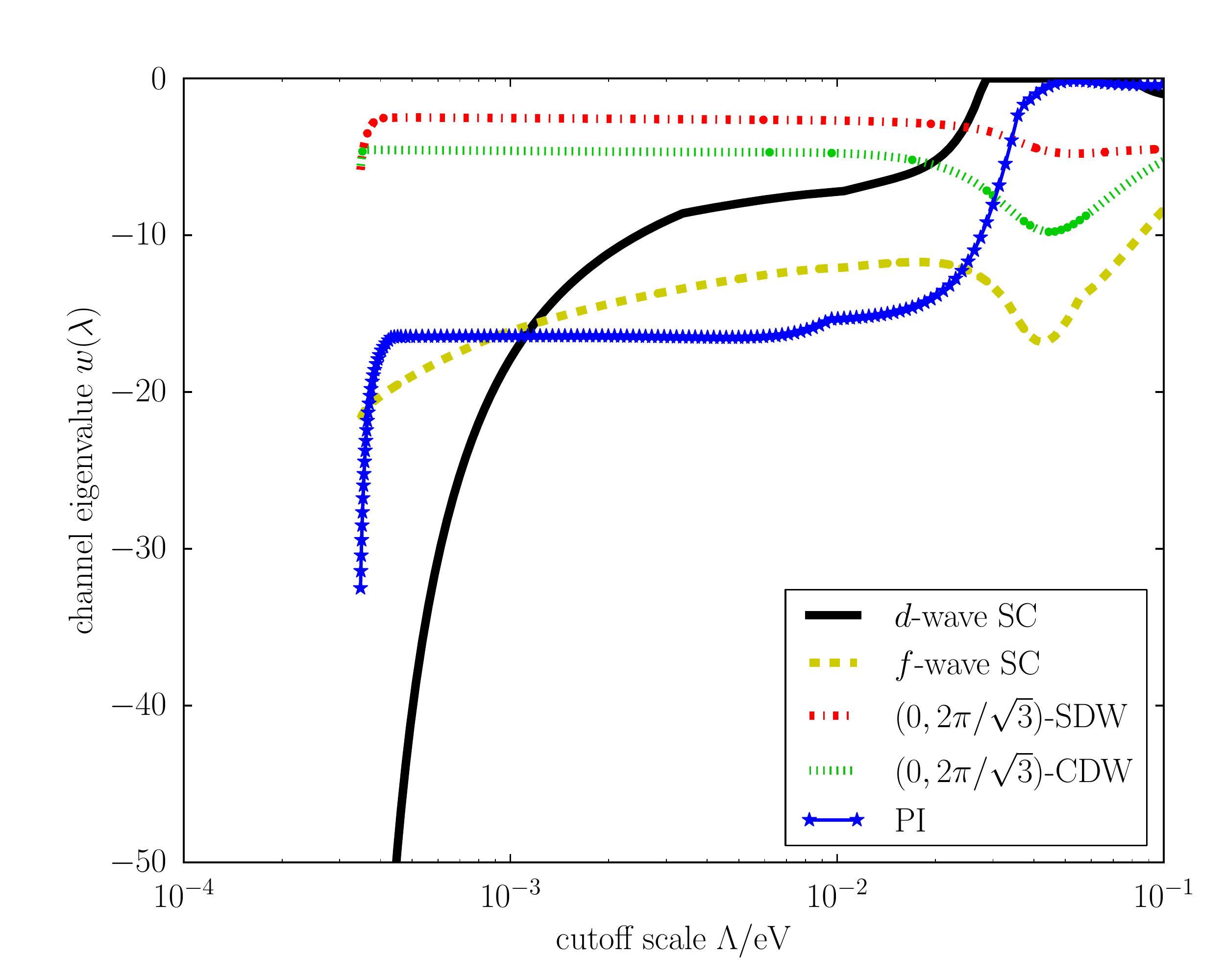}
  \caption{Eigenvalue flow of several RG channels for $g_1=g_2=0.4$eV and $g_3=g_4=1.5$eV. The first
    diverging mode is the doubly-degenerate $d$-wave superconducting (SC) channel,
    followed by the subleading doubly-degenerate $d$-wave Pomeranchuk channel (PI). Due to imperfect nesting, neither the charge density wave (CDW) nor the spin-density wave (SDW) play a significant role.
    }
  \label{fig:flow}
\end{figure}

\begin{figure}[t]
  \centering
  \includegraphics[width=8cm]{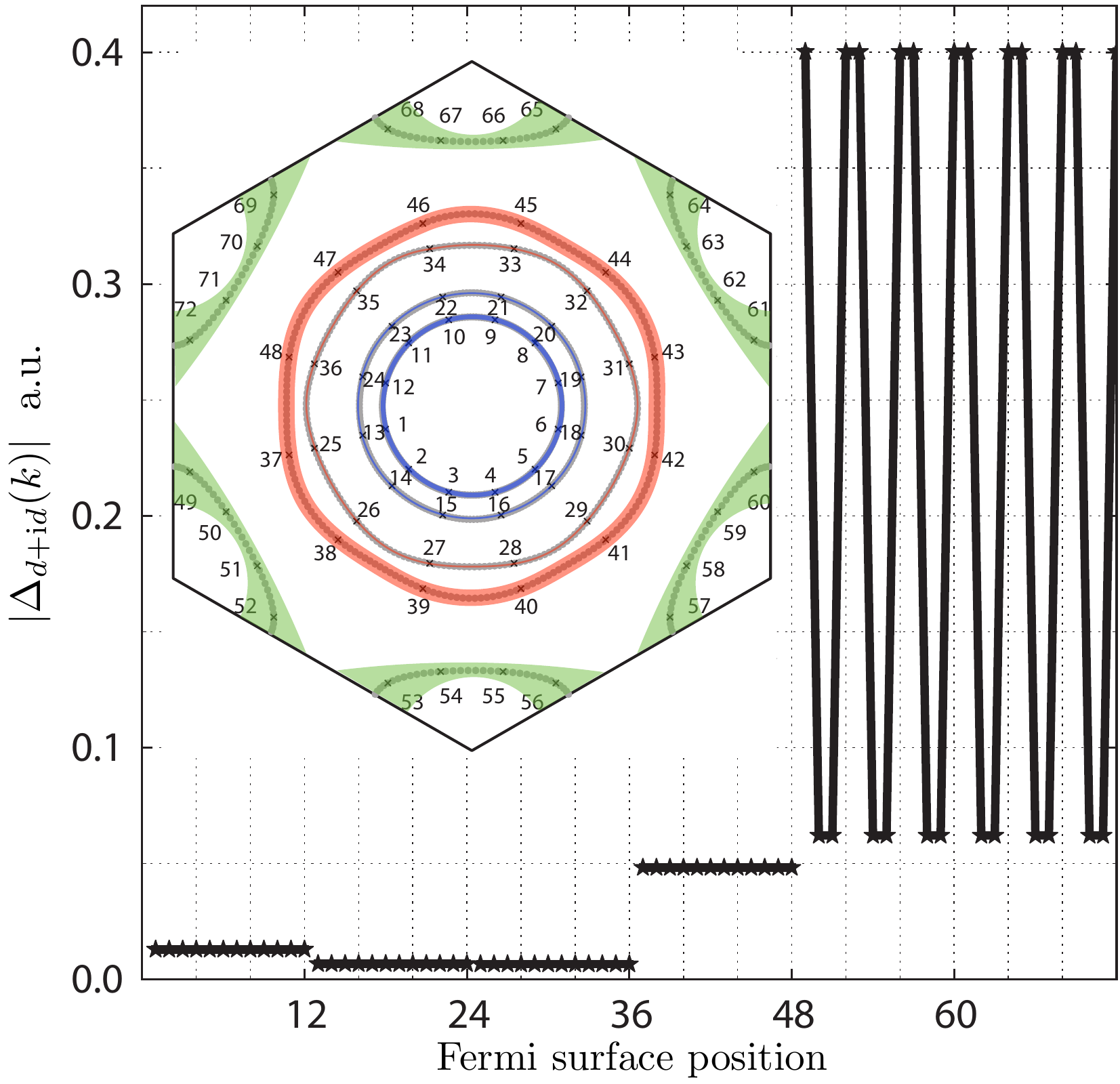}
  \caption{Representative $d+id$-gap structure plotted along the discretized Fermi
    pockets ($g_1=g_2=0.4$eV and $g_3=g_4=1.5$eV). There is significant gap anisotropy on the pockets around $K$ and $K'$. The pockets centered around
    $\Gamma$ only show small, isotropic gaps mainly induced by the proximity-coupling
    to the outer pockets. The inset visualizes the gap magnitude and shows the discretization of the Fermi surfaces.}
  \label{fig:gap}
\end{figure}

\emph{Renormalization group results.}
Using FRG,
we compute the effective, renormalized interaction described by the 4-point
function (4PF)
$V_{\Lambda}(\bs{k}_1,\alpha;\bs{k}_2,\alpha';\bs{k}_3,\beta;\bs{k}_4,\beta')\psi_{\bs{k}_4\beta's}^{\dagger}\psi_{\bs{k}_3\beta\bar{s}}^{\dagger}\psi_{\bs{k}_2\alpha's}^{\phantom{\dagger}}\psi_{\bs{k}_1\alpha\bar{s}}^{\phantom{\dagger}},$
where the flow parameter is the IR cutoff $\Lambda$ approaching the
Fermi surface, and with $\bs{k}_{1}$ to $\bs{k}_{4}$ the incoming and
outgoing momenta. The starting conditions
are given by the bandwidth serving as a UV cutoff, with the bare
initial interactions in Eq.~\eqref{eq:Hint} serving as the initial 4PF. The diverging channels of the 4PF
under the flow to the Fermi surface signal the nature of the
instability, which in our case we find to be located in the Cooper channel. 
The 4PF in the Cooper channel can be decomposed into
different eigenmode contributions
\begin{equation}
V^{\text{SC}}_{\Lambda} (\bs{k},-\bs{k},\bs{p})= \sum_i c_i^{\text{SC}}(\Lambda) f^{\text{SC},i}(\bs{k})^* f^{\text{SC},i}(\bs{p}),
\label{decomp}
\end{equation} 
where $i$ is a summation index over all eigenvalues, 
whose number equals the number of
discretized momentum points along the Fermi surfaces (inset Fig.~\ref{fig:gap}).
The leading instability of that channel corresponds to the $c_1^{\text{SC}}(\Lambda)$ that first diverges under the
flow of $\Lambda$ (Fig.~\ref{fig:flow}).

At a larger cutoff scale, the first significant feature of the channel
flow is a bump in several channels due to the van-Hove
singularity located at this cutoff distance from the Fermi level. This
is accompanied by an enhanced spin-triplet pairing interaction (see Fig.~\ref{fig:flow}). At
a smaller cutoff scale, a ($d$-wave) Pomeranchuck channel starts to grow and, along
with it, the pairing in the $d$-wave channel. Finally, at low scales,
the $d$-wave channel diverges. The initial switch from triplet to
singlet in the Cooper channel under the renormalization group flow is common in specific multiband
fermiologies with significant unnested density of states at lower
energies, seen in LiFeAs~\cite{PhysRevB.84.235121} to name an example. As a more
peculiar feature, however, note how the subleading Pomeranchuck
and the leading
superconducting eigenvalue appear to show significant interdependence, implicating that fluctuations in the Pomeranchuck channel
might contribute to seeding superconductivity. 
 
Figure~\ref{fig:gap} shows the ($d+\mathrm{i}d$) gap structure resulting from the $d$-wave superconducting
instability. The gap shows a significant anisotropy along the outer pockets due to higher harmonic
contributions in the $d$-wave form factors. In addition, there is a clear
difference of the gap scale on the inner bands around the $\Gamma$
point and the outer bands around the $K$, $K'$ points, with the largest gap at the $\kk$ points closest to the van-Hove singularity. This again
emphasizes the importance of the outer band due to its proximity to
the van-Hove singularity. Nevertheless, inclusion of all
bands is crucial for obtaining the $d$-wave  order parameter. If it were
only for the outermost bands, previous calculations suggest an
$f$-wave order parameter to be dominant~\cite{raghu:2010b,
  kiesel:2012}. However, such a state would have nodes on the
$\Gamma$-centered pockets, and therefore necessarily loses
condensation energy as compared to the dominant $d$-wave solution. 

\begin{figure*}[t]
  \begin{center}
    \includegraphics[width=\textwidth]{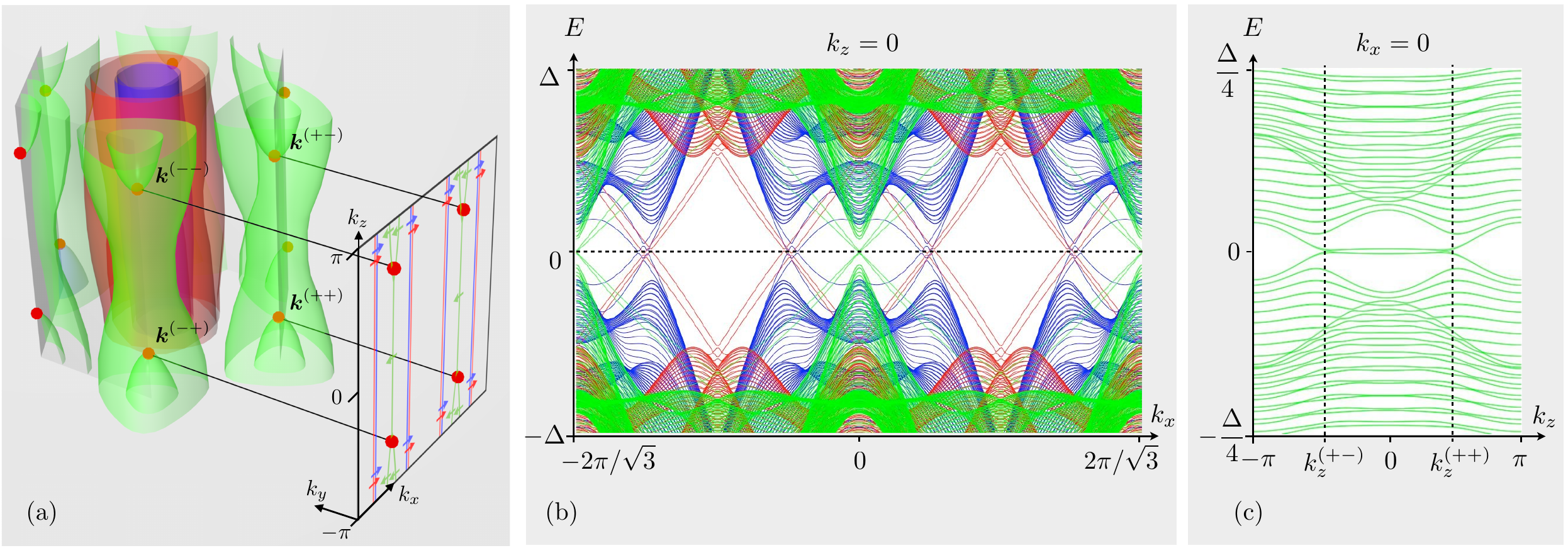}
  \end{center}
  \caption{(Color online)
  Fermi surface structure and edge states of the $E_g$ superconducting state of SrPtAs. a) Bulk Fermi surfaces with Majorana-Weyl nodes of the superconducting order parameter (red dots) and their projection to the surface BZ of the (010) surface. In the surface BZ, the Fermi surfaces and Fermi arcs of the chiral Majorana fermions are sketched, with the arrows indicating their group velocity. b) Cut along $k_z=0$ of the spectrum of a 150 layer slab stacked along the (010) direction showing Majorana fermion surface states. c) The cut along $k_x=0$ of the spectrum of a 400 layer slab stacked along the (010) direction shows the position of two Majorana-Weyl points as well as the chiral Majorana surface modes connecting them.
  }
  \label{fig:edge states}
\end{figure*}
\emph{Characterization of the chiral $d$-wave state.} 
Having identified the $d$-wave superconducting state as the dominant instability using FRG, we now discuss its momentum-space structure and its topological properties in more detail.
We first note that the spin-orbit-coupling term $\alpha^{(b)}_{\bs k}$ [Eq.~\eqref{eq:bands}] of Ref.~\cite{youn:2012} allows in principle for an admixture of a small (staggered) chiral $p$-wave component to the dominant chiral $d$-wave state~\cite{goryo:2012}. We thus consider here the more general form of the gap function  
\begin{equation}
\Delta^{\pm}_{\bs{k},l}
=
\mathrm{i}\sigma_2\,\Delta^\pm[e^\pm_{\bs{k}}\sigma_0+r(-1)^l o^\pm_{\bs{k}}\sigma_3],
\label{eq: gap function}
\end{equation}
where $l=1,2$ denotes the layer index, 
$e^+_{\bs{k}}=(e^-_{\bs{k}})^*=\sum_n w^n \cos \bs{k}\cdot \bs{T}_n$, and
$o^+_{\bs{k}}=(o^-_{\bs{k}})^*=\sum_n w^n \sin \bs{k}\cdot \bs{T}_n$, with $w^n=\exp (\mathrm{i}2\pi n/3),\ n=1,2,3$.
The Pauli matrices $\sigma_2$, $\sigma_3$ and the $2\times2$ unit matrix $\sigma_0$ act in spin space.

The gap function $\Delta^{\pm}_{\bs{k},l}$ preserves the $s_z$ spin-rotation symmetry, but breaks time-reversal symmetry by spontaneously choosing one of the two {chiralities `$\pm$'~\cite{goryo:2012}}. It has three nodal lines in the BZ which run parallel to the $k_z$-axis, one at $k_x=k_y=0$ and one at each corner of the hexagonal basal plane of the BZ at $k_x=0$, $k_y=\pm4\pi/3$.
These nodal lines do not intersect with the two pairs of Fermi surfaces that are centered around the $\Gamma$ point [blue and red Fermi surfaces in Fig.~\ref{fig:edge states} (a)]. They do, however intersect with the cigar-shaped sheet of the pair of Fermi surfaces that are centered around the BZ corners
[green Fermi surfaces in Fig.~\ref{fig:edge states} (a)]. 
The low energy effective theory of the Bogoliubov quasiparticles near these four intersection points 
\begin{equation}
\bs{k}^{(\lambda,\lambda')}
=
\left(
0,\lambda\frac{4\pi}{3},\lambda'\arccos\frac{3\sqrt{3}\alpha^{(3)}+2\mu^{(3)}+3t^{(3)}}{2t^{\prime(3)}_{z}}
\right)^{\mathsf{T}},
\label{eq: Weyl points}
\end{equation}
with $\lambda=\pm$, $\lambda'=\pm$,
is that of a three-dimensional Majorana-Weyl fermion~\cite{conyers:1937, volovik:2003, xiangang:2011,1367-2630-15-6-065001, meng:2012} with a momentum-linear dispersion.
Thus, the chiral $d$-wave state of SrPtAs is a Weyl superconductor.

If the $s_z$ spin rotation symmetry is preserved, the system belongs to symmetry class A in the classification of Ref.~\cite{schnyder:2008}. Class A is trivial in 3D, but admits a $\mathbb{Z}$ classification in 2D, with the Chern number of the Bogoliubov bands as the associated topological invariant. If translational symmetry holds, the Chern number can be defined for a Bloch Bogoliubov-de Gennes Hamiltonian in any 2D plane in the BZ that does not intersect the Majorana-Weyl nodes $\bs{k}^{(\lambda,\lambda')}$. As SrPtAs is a layered material stacked in the $z$-direction, it is natural to compute the Chern number in the $k_x$-$k_y$-plane as a function of $k_z$. We obtain for the fully gapped bands $b=1,2$ that $C^{(b)}_{k_z}=8$ independent of $k_z$ and
\begin{equation}
C^{(3)}_{k_z}=
\begin{cases}
-4&\text{for}\ |k_z|<|k^{(+,+)}_{z}|,\\
-8&\text{for}\ \pi>|k_z|>|k^{(+,+)}_{z}|
\end{cases}
\end{equation}
for the band with the Majorana-Weyl nodes.

Finally, we discuss consequences of the nonzero Chern numbers: Planes perpendicular to the [001] direction support (one-dimensional) topological surface states, that are chiral Majorana fermions. Figure~\ref{fig:edge states}(a) sketches the Fermi  surfaces associated with the (2D) surface states [see Fig. 4(b)], which in 3D live in planes parallel to [001]. In particular, a pair of Fermi arcs spans between the projections in the surface BZ of the Majorana-Weyl points $\bs{k}^{(+,+)}$ and $\bs{k}^{(+,-)}$ as well as between the projections of $\bs{k}^{(-,+)}$ and $\bs{k}^{(-,-)}$ [see Fig.~\ref{fig:edge states}(c)].
The topological response associated with these surface states is a spontaneous thermal Hall effect. The reduced thermal Hall conductivity $\kappa_{xy}/T$ takes universal quantized values for fully gapped 2D superconductors. In 3D, however, it carries extra dimension of inverse length and therefore depends explicitly on the height of the unit cell $c$ in $z$-direction and the separation $2|k^{(+,+)}_z|$ between the Majorana-Weyl points
\begin{equation}
\kappa_{xy}=
\frac{k^2_{\mathrm{B}}T}{24} 
\int_0^{2\pi}\frac{\mathrm{d}k_z}{c}\sum_{b}C^{(b)}_{k_z}
=
\frac{k^2_{\mathrm{B}}T}{3 c} 
\,\left(2\pi+ |k^{(+,+)}_z|\right).
\end{equation}

\emph{Conclusion.}
Using FRG, we provide strong evidence that superconductivity in SrPtAs realizes a chiral $d$-wave state. Given the hexagonal symmetry, we identify the specific multiband fermiology and its Fermi pockets in proximity to van Hove singularities as the main ingredients that stabilize this pairing state.  
The obtained state is fully consistent with the existing experimental data. First,  $\mu$SR measurements constrain the superconducting state to break TRS and to have no line nodes on the Fermi surfaces.
Second, recent nuclear magnetic resonance measurements found evidence for multi-gap superconductivity and a suppressed coherence peak that is consistent with a chiral $d$-wave order parameter~\cite{brueckner:2013}. For a quantitative prediction of the $d$-wave anisotropy, as well as the relative gap sizes on the different bands, a more refined, orbital-resolved  description of the interacting Hamiltonian is necessary and is beyond the scope of this paper.

The chiral $d$-wave order parameter is \emph{energetically} explained by resorting to a strictly two-dimensional approximation of the band structure of SrPtAs. However, the weak three-dimensionality of SrPtAs has important consequences for its \emph{topological} properties: We find SrPtAs to be a superconductor with protected Majorana-Weyl nodes in the bulk and (Majorana) Fermi arcs on the surface, along with other topological Majorana surface states. For an experimental investigation of the exotic surface properties, including the thermal Hall response, and directional anisotropies of the chiral $d$-wave state, the availability of high-quality single crystals will be crucial. 

\emph{Acknowledgements.} 
MHF acknowledges support from the NSF grant no. DMR-0955822, from NSF grant no. DMR-1120296 to the Cornell Center for Materials Research, and the Swiss Society of Friends of the Weizmann Institute of Science.
CP, WH, and RT are supported by DFG-FOR 1458/2. RT is supported by the European Research Council through ERC-StG-2013-Thomale-336012. TN acknowledges financial support from the Swiss National Science Foundation.
\bibliography{ref}
\end{document}